\documentclass{aa}  

\usepackage{graphicx}
\usepackage{txfonts}
\usepackage{hyperref}
\usepackage{xcolor}

\newcommand{\planck}{\textsl{Planck}}

\newcommand{\rosat}{\textsl{ROSAT}}
\newcommand{\chandra}{\textsl{Chandra}}
\newcommand{\xmm}{\textsl{XMM-Newton}}

\newcommand{\srg}{\textsl{SRG/eROSITA}} 
\newcommand{\erosita}{\textsl{eROSITA}} 

\newcommand{\beq}{\begin{equation}}
\newcommand{\eeq}{\end{equation}}
\newcommand{\beqa}{\begin{eqnarray}}
\newcommand{\eeqa}{\end{eqnarray}}

\newcommand{\ie}{i.e.\xspace}

\def\msun{{\rm M}_{\odot}}

\def\zsun{Z_{\odot}}

\begin{document} 

    \title{
     X-ray emission from cosmic web filaments in \srg{} data}
     
    \author{H. Tanimura\inst{1,2} \and N. Aghanim\inst{1} \and M. Douspis\inst{1} \and N. Malavasi\inst{3}}

    \institute{
    Universit\'{e} Paris-Saclay, CNRS, Institut d'Astrophysique Spatiale, B\^atiment 121, 91405 Orsay, France \and
    Kavli IPMU (WPI), UTIAS, The University of Tokyo, Kashiwa, Chiba 277-8583, Japan \and
    Faculty of Physics, Ludwig-Maximilians-Universit\"{a}t, Scheinerstr, Munich, 81679, Germany \\
    \email{hideki.tanimura@ias.u-psud.fr} }

    \date{}

    \abstract 
    {Using the publicly available \erosita\ Final Equatorial Depth Survey (eFEDS) data, we detected the stacked X-ray emissions at the position of 463 filaments at a significance of 3.8$\sigma$ based on the combination of all energy bands. In parallel, we found that the probability of the measurement under the null hypothesis is $\sim$0.0017. The filaments were identified with galaxies in the Sloan Digital Sky Survey survey, ranging from 30 Mpc to 100 Mpc in length at 0.2 < $z$ < 0.6. The stacking of the filaments was performed with the eFEDS X-ray count-rate maps in the energy range between 0.4 -- 2.3 keV after masking the resolved galaxy groups and clusters and the identified X-ray point sources from the \rosat, \chandra, \xmm, and \erosita\ observations. In addition, diffuse X-ray foreground and background emissions or any residual contribution were removed by subtracting the signal in the region between 10 -- 20 Mpc from the filament spines. 
    For the stacked signal, we performed an X-ray spectral analysis, which indicated that the signal is associated with a thermal emission. According to a model with the astrophysical plasma emission code for the plasma emission and with a $\beta$-model gas distribution with $\beta$=2/3, the detected X-ray signal can be interpreted as emission from hot gas in the filaments with an average gas temperature of $1.0^{+0.3}_{-0.2}$ keV and a gas overdensity of $21\pm5$ at the center of the filaments. 
    }
    \keywords{cosmology: observations -- large-scale structure of Universe -- diffuse radiation, X-rays: diffuse background}

    \maketitle
%
\section{Introduction}
\label{sec:intro}

In the standard $\Lambda$CDM cosmology, more than $\sim$95\% of the energy density in the Universe is in the form of dark matter and dark energy, whereas baryonic matter only comprises $\sim$5\% \citep{Planck2020VI}. In this model, the structures form via the gravitational collapse of matter in the expanding Universe (e.g., \citealt{Zeldovich1982}) and cause the large-scale structure of the Universe to be organized in a web-like pattern that is called the cosmic web \citep{Bond1996}.

The structure formation has been widely studied with numerical simulations. The simulations illustrated that the cosmic web consists of nodes, filaments, sheets, and voids (\ie, \citealt{Aragon2010, Aragon2010b, Cautun2013, Cautun2014}). Nodes are dense regions that are interconnected by filaments and sheets, voids are regions of extremely low density that occupy most of the volume in the Universe, and the matter density currently extends over six orders of magnitude \citep{Cautun2014}. In this variety of cosmic environments, baryons are further influenced by radiative cooling and energetic feedback from star formation, supernovae, and active galactic nuclei (AGN), evolving in a complicated manner \citep{Cen2006, Martizzi2019, Daniela2022}. For example, at high redshifts ($z \gtrsim 2$), most of the expected baryons are found in the Ly$\alpha$ absorption forest: the diffuse, photoionized intergalactic medium (IGM) with a temperature of $10^4$--$10^5$ K (e.g., \citealt{Weinberg1997, Rauch1997}). However, at redshifts $z \lesssim 2$, the observed baryons in stars, the cold interstellar medium, residual Ly$\alpha$ forest gas, ionized oxygen (OVI), broad Ly$\alpha$ absorbers (BLAs), and hot gas in clusters of galaxies account for $\sim$70\% of the expected baryons; the remainder has yet to be identified (e.g., \citealt{Fukugita2004, Shull2012}) and constitutes the so-called missing baryons. 

Hydrodynamical simulations suggest that 40--50\% of the baryons might be in the form of shock-heated gas in the cosmic web filaments between clusters of galaxies \citep{Cen2006}. This warm hot intergalactic medium (WHIM) with temperatures ranging between $10^5$ and $10^7$ K is considered to account for most of the missing baryon \citep{Shull2012}. Therefore, extensive searches for the WHIM have been performed in the far-ultraviolet domain, but also in the X-rays and based on the thermal Sunyaev-Zel'dovich (tSZ) effect \citep{Sunyaev1970, Sunyaev1972} (e.g., \citealt{Fujita1996, Fujita2008, Tittley2001, Dietrich2005, Werner2008, Planck2013IR-VIII, Sugawara2017, Alvarez2018, Bonjean2018, Nicastro2018, Kovacs2019, deGraaff2019, Tanimura2019a, Tanimura2019b, Tanimura2020y, Tanimura2020x, Hincks2022}). However, the WHIM is difficult to observe because the signal is relatively weak and the morphology of the filaments is complex. 

The filamentary structure is indeed hard to identify because its density is low and the morphology is complex. Therefore, several algorithms have been developed to detect them, such as Spineweb \citep{Aragon2010b}, DisPerSE \citep{Sousbie2011, Sousbie2011b}, NEXUS/NEXUS+ \citep{Cautun2013}, Bisous \citep{Tempel2014}, and T-ReX \citep{Bonnaire2020, Bonnaire2022}. 
These algorithms have been applied to observational data and identified the filamentary pattern, such as in the distribution of galaxies from the Sloan Digital Sky Survey \citep[SDSS]{Gunn2006}  (e.g., \citealt{Tempel2014, Chen2015, Chen2016, Malavasi2020, Carron2022}), the Cosmological Evolution Survey \citep[COSMOS]{Scoville2007}  (e.g., \citealt{Laigle2018}), and the VIMOS public extragalactic redshift survey \citep[VIPERS]{Scodeggio2018} (e.g., \citealt{Malavasi2017, Moutard2016b, Moutard2016}). The properties of the filaments were then studied at different wavelengths.
For example, \cite{Tempel2014} found that the filaments contain 35 -- 40\% of the total galaxy luminosity. \cite{Chen2016} showed that galaxies close to filaments are generally brighter than those at significant distance from filaments. \cite{Laigle2018} found that passive galaxies are more confined in the core of the filament than star-forming galaxies. 
\cite{Bonjean2020} unveiled the galaxy distribution and their properties in the SDSS filaments, such as star formation rate, galaxy types, and stellar mass. \cite{Tanimura2020y} (hereafter T20y) detected the WHIM gas at a significance of $\sim 4\,\sigma$ with the tSZ effect and at a significance of $\sim 8\,\sigma$ in the CMB lensing signal and measured both the gas density and temperature. 

In addition, more recently, \cite{Tanimura2020x} (hereafter T20x) detected the X-ray emission from the WHIM gas for the first time in large cosmic filaments at a significance of $\sim 4\,\sigma$ using the \rosat\ data\footnote{https://www.jb.man.ac.uk/research/cosmos/rosat/} provided by \cite{Snowden1994} and \cite{Snowden1997}. The authors constrained both the gas temperature of $T \sim 0.9^{+1.0}_{-0.6}$ keV and its overdensity of $\delta \sim 30 \pm 15$ at the center of the filaments. They further forecast the detection of the X-ray emission from filaments in the \srg\ data assuming different hypotheses for the gas densities and temperatures. An actual observational study with \srg{} was performed for the Abell 3391/95 galaxy cluster system in \cite{Reiprich2021}, and the authors discovered evidence of X-ray emission from warm gas in the bridge region, which extends over 15 Mpc.

As an extension of T20x work, we use the publicly available \srg\ data in this study to assess an X-ray emission associated with large cosmic filaments. Based on this, we estimate the gas density and temperature in the filaments. The paper is organized as follows: Section \ref{sec:data} summarizes the datasets we used in our analyses. Section \ref{sec:ana} explains the stacking method for measuring the X-ray emission from filaments. Section \ref{sec:interp} presents the model we used to interpret our measurements. The possible systematic uncertainties in our measurements are discussed in Section \ref{sec:systematics}. We conclude the paper with a discussion and conclusions in Section \ref{sec:discussion} and Section \ref{sec:conclusion}. Throughout this work, we adopt the $\Lambda$CDM cosmology from \cite{Planck2020VI} with $\Omega_{\rm m} = 0.3158$, $\Omega_{\rm b} = 0.0494$, and $H_0 = 67.32$ km s$^{-1}$ Mpc$^{-1}$. All masses are quoted in solar mass, and $M_{\Delta}$ is the mass enclosed within a sphere of radius $R_{\Delta}$ such that the enclosed density is $\Delta$ times the critical density at redshift $z$. Uncertainties are given at the 1$\sigma$ confidence level.

\section{Datasets}
\label{sec:data}

\subsection{eROSITA Final Equatorial Depth Survey}
\label{subsec:efeds}
The extended ROentgen Survey with an Imaging Telescope Array (\erosita) \citep{predehl2010} is achieving full-sky X-ray surveys with its instrument on board the Russian Spektrum-Roentgen-Gamma satellite\footnote{http://hea.iki.rssi.ru/SRG} (SRG). After completing four years of observation, \erosita\ will achieve eight full-sky surveys, resulting in a sky-averaged exposure time of $\sim$2 ks. This will allow the detection of about $100\,000$ galaxy clusters \citep[SRG/eROSITA Science Book:][]{Merloni2012}. 

In this analysis, we use the \erosita\ Final Equatorial Depth Survey\footnote{https://erosita.mpe.mpg.de/edr/eROSITAObservations/ \label{foot:efeds}} (eFEDS) \citep{brunner2022}. It has been carried out as a part of the calibration and performance verification (Cal-PV) program between mid-September and mid-December 2019 (about 360 ks, or 100 hours, in total) with all the seven \srg\ Telescope Modules (TM 1-7) in operation. The telescope modules performed scanning observations of $\sim$140 square degrees, composed of four individual rectangular raster-scan fields of $\sim$35 square degrees each. However, due to an unrecognized malfunction of the camera electronics, 28\% of the TM6 data of eFEDS field I and 48\% and 43\% of the TM5 and TM6 data, respectively, of eFEDS field II could not be used, resulting in a reduced exposure depth in the affected areas of up to $\sim$30\%. Because the data in these fields have higher noise than other eFEDS fields, we discarded the data in these fields from our analysis. 

A set of calibrated data products from \erosita, the Early Data Release (EDR)\ref{foot:efeds}, is made public. They consist of calibrated event files with energies between 0.2 - 10 keV and image, exposure, and background maps in the energy range of 0.2 - 0.6 keV, 0.6 - 2.3 keV, and 0.2 - 2.3 keV. EDR also includes the X-ray sources detected at 0.2 - 2.3 keV with a detection likelihood threshold of five. The sources with a detection likelihood higher than six are selected as a main eFEDS catalog (27910 main sources), and the sources with a detection likelihood between five and six are selected as a supplementary eFEDS catalog (4774 supplementary sources). EDR also provides a catalog of 542 galaxy group and cluster candidates, including their position, temperature, luminosity, and flux \citep{Liu2022}. \cite{Chiu2022} estimated the cluster mass calibrated with the weak-lensing data from the Hyper Suprime-Cam (HSC) Subaru Strategic Program survey \citep{Aihara2018}. We here aim at detecting and characterizing the X-ray signal in filaments, and all these point-like and extended sources are therefore masked in our analysis. 

\subsection{Filament catalog from SDSS}
\label{subsec:filcat}
\begin{figure}
    \centering
    \includegraphics[width=0.49\linewidth]{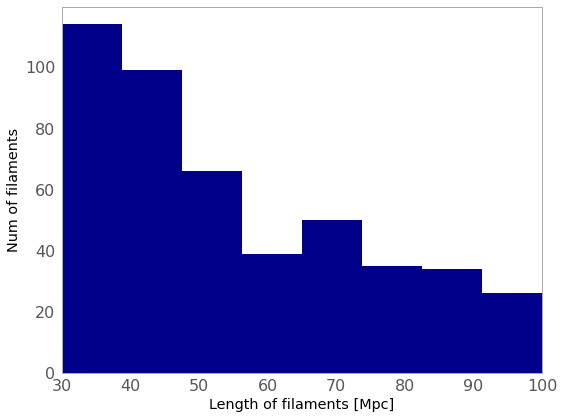} 
    \includegraphics[width=0.49\linewidth]{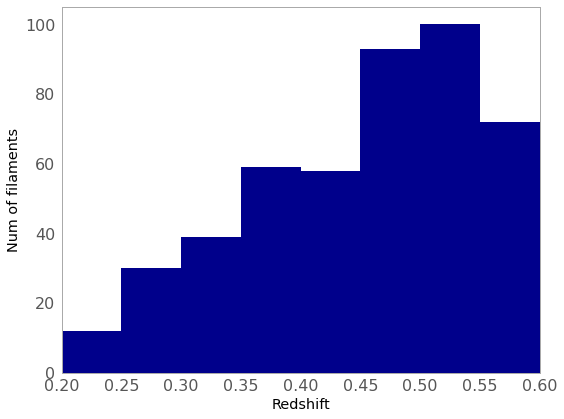} 
    \caption{Length (left) and redshift (right) distribution of the 463 filaments used in our analysis. }
    \label{fig:hist-fil}
    \end{figure}

We used the filament catalog constructed in \cite{Malavasi2020} by applying the discrete persistent structure extractor (DisPerSE) algorithm \citep{Sousbie2011} to the LOWZ and CMASS spectroscopic galaxies in the Baryon Oscillation Spectroscopic Survey (BOSS) \citep{Reid2016}. The DisPerSE method computes a gradient of a density field and identifies a critical point at which the gradient is zero. The critical points can then be classified as maxima of the density field (e.g., groups or clusters of galaxies), minima of the density field (associated with voids of galaxies), and saddles (local density minima bounded to structures, e.g., filaments). DisPerSE then defines filaments as field lines with a constant gradient that connect critical points (maxima and saddles). \cite{Malavasi2020} reported a total of 63,391 filaments, 601 of which overlap with the eFEDS footprint. Their lengths range from 30 to 100 Mpc at $0.2<z<0.6$ (in order to focus on large cosmic filaments and not on short bridges connecting pairs of clusters; see \cite{Daniela2021} for the gas distinction between these two categories). We finally used 463 of these 601 filaments in our analysis (see Sect. \ref{sec:ana}). The length and redshift distribution of the 463 filaments are shown in Fig.~\ref{fig:hist-fil}. The angular size of the filament lengths extends from 0.8 to 6.7 degrees on the sky.

\subsection{Point-source and cluster catalogs}
\label{subsec:extcat}
In order to remove the contribution from the foreground and background galaxy groups and clusters, we masked the objects from the catalogs listed in T20x. These catalogs include 1653 tSZ clusters from the \planck\ observations \citep{Planck2016XXVII} and 1743 MCXC X-ray clusters from the \rosat\ X-ray survey \citep{Piffaretti2011}, as well as 26\,111 redMaPPer clusters \citep{Rykoff2014}, 158\,103 WHL clusters \citep{Wen2012, Wen2015}, and 46\,479 AMF clusters \citep{Banerjee2018} detected from the galaxies in the Sloan Digital Sky Survey (SDSS) survey. The mass and redshift distributions of these objects are shown in \cite{Tanimura2019a}, and the mass distribution in the union catalog reaches down to $\sim 3 \times 10^{13} \, \msun$. We masked all these groups and clusters of galaxies from the union catalog that fall in the eFEDS footprint.
In addition, as mentioned in Sect. \ref{subsec:efeds}, we masked the 542 galaxy clusters that are directly detected in the eFEDS observation. 

Moreover, we masked the X-ray point sources detected in the \rosat, \chandra, and \xmm\ observations. They include 135\,118 point-like sources in the Second \rosat\ All-Sky Survey Point Source Catalog (2RXS)\footnote{http://www.mpe.mpg.de/ROSAT/2RXS}\citep{boller2016}, 317\,167 X-ray sources in the \chandra\ Source Catalog (CSC 2.0)\footnote{http://cxc.cfa.harvard.edu/csc2/}\citep{evans2010}, and 775\,153 sources in the third-generation catalog of X-ray sources from the \xmm\ observatory (3XMM-DR8)\footnote{https://www.cosmos.esa.int/web/xmm-newton/xsa\#download}\citep{rosen2016}. Here again, we masked all the X-ray point sources that fall in the eFEDS footprint, in addition to those detected in EDR.

\section{Stacking analysis}
\label{sec:ana}

This section describes the procedure by which we stacked the eFEDS maps at the position of the cosmic filaments. We then estimate the X-ray signal associated with the filaments. 
For the stacking method, we followed the procedure initiated in T20y and optimized for the X-ray data in T20x. 

\subsection{Masking clusters and point sources}
\label{subsec:mask}

Before stacking the X-ray eFEDS maps, we masked out all the sources listed in Sect. \ref{sec:data}. 
The point sources were masked by a radius of 30 arcsec, which is much larger than the angular resolution of  \srg,\  which has an on-axis angular resolution of 5.7 -- 9.3 arcsec in full width at half maximum (FWHM), depending on the detector (TM 1--7) \citep{Predehl2021}. The average residual X-ray flux from masked point sources is thus $\sim$10\% at most. The galaxy groups and clusters were masked by $3 \times R_{500}$, corresponding to about 1.5 times their virial radius. In addition, we masked the maxima of the galaxy density field identified by DisPerSE, which are possible locations of unresolved galaxy clusters. We masked these maxima by a radius of 30 arcsec. We did not mask other critical points identified by DisPerSE, such as minima and saddles, because they are associated with low-density regions. We checked that our results do not change by varying these mask sizes, as shown later in Sect. \ref{sec:systematics}. In our analysis, we study the excess signal relative to the background, hence the mean signal in the background region was subtracted for each filament (see Sect. \ref{subsec:method}). This background-subtraction procedure further removes the residual contribution from masked point sources and groups and clusters of galaxies.
As a result of this masking procedure and of the removal of the low-exposure eFEDS fields, in which the noise is higher (see Sect. \ref{sec:data}), the available eFEDS area for our stacking analysis corresponds to $\sim$ 32\% of the initial eFEDS observed area.
    
\subsection{Stacking method}
\label{subsec:method}

We used the eFEDS image and exposure maps at 0.2 - 0.6 keV and 0.6 - 2.3 keV provided in EDR. We computed the count rate of the X-ray photons in the eFEDS field and stacked the count-rate map at the position of the filaments. We did not use the provided eFEDS background maps in our analysis because our analysis pipeline includes a background-subtraction step. We checked that subtracting the background maps from the image maps did not change our results.

\begin{figure}
    \centering
    \includegraphics[width=\linewidth]{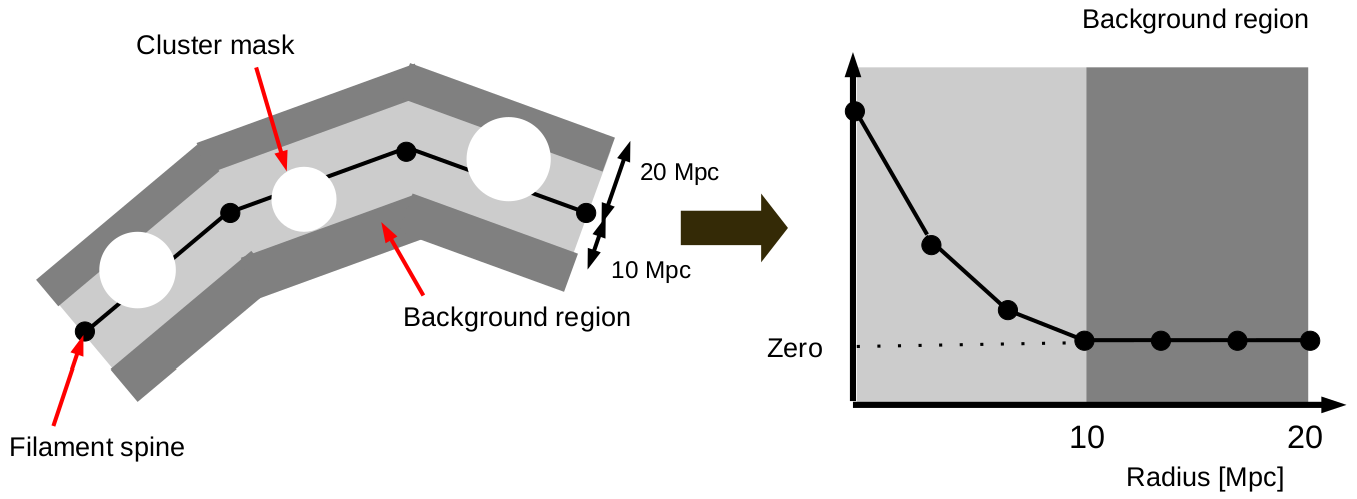} 
    \caption{Computation of a radial profile of one filament. In the left panel, black dots show the extremities of the segments of filament given by DisPerSE. The straight black line connecting these extremities constitutes the filament spine. The area within a physical distance up to 20 Mpc from the filament spine is shown in light and dark gray. This area was used to compute a radial profile of the filament, as shown in the right panel. The area within a physical distance of 10--20 Mpc from the filament spine is shown in dark gray. It is defined as a background region. The average signal in the background region is subtracted from the radial profile. White disks are masked regions that were excluded from our analysis.}
    \label{fig:anafil}
    \end{figure}

Then, we followed the stacking method in T20x, which is illustrated in Fig.~\ref{fig:anafil}. 
(i) In the left panel of Fig.~\ref{fig:anafil}, the black dots represent the extremities of the segments, defining the filament by DisPerSE. The black straight line connecting these extremities constitutes the filament spine. 
We used data within 20 Mpc from the filament spine. For each data point within the 20 Mpc region, the closest distance to the filament spine was computed. Then, the data were divided into eight distance bins up to 20 Mpc, as illustrated in Fig.~\ref{fig:xprof}, and the average value of the count rate in each distance bin was computed. 
From the extent of the data profile in Fig.~\ref{fig:xprof}, data within 10 Mpc from the filament spine were used to estimate the signal-to-noise ratio (S/N) of our measurement, shown in light gray, and those between 10 -- 20 Mpc were used for our background estimate, described below in the step (iii), shown in dark gray. 

(ii) In the left panel of Fig.~\ref{fig:anafil}, white disks are the masked regions; data in theses regions were not used in our analysis. A filament may be largely masked if it is located close to our excluded area or massive clusters in the foreground or background (see Sect. \ref{subsec:mask}). In this extreme situation, we discarded the filament when the resulting radial profile had empty radial bins (i.e., no data were accumulated in the bins because of the masking procedure). This process reduced the number of SDSS filaments in the eFEDS region from 601 to 473 at 0.2 - 0.6 keV and to 463 at 0.6 - 2.3 keV. (Two effects cause the lower number of available filaments at 0.6 - 2.3 keV: empty bins due to the masking procedure, and empty bins due to the lower number of X-ray photons at higher energies than those at lower energies.) In our analysis, we used the commonly selected 463 filaments.

(iii) To estimate the excess of the X-ray signal associated with the filament, we subtracted the average background signal in the outskirts in a region covering 10--20 Mpc from the filament spine that is defined as the local background signal. It corresponds to the dark gray area in the left panel in Fig.~\ref{fig:anafil}. This subtraction mitigates the contamination from both the foreground and background emissions, such as the diffuse Galactic and extragalactic emissions, as well as residual signal from point sources or systematic effects.

(iv) Steps (i)--(iii) were repeated for all the usable filaments, and the ensemble of background-subtracted radial profiles was obtained. To account for the mask effect that does not impact each filament in the same manner, we computed the weighted average of the background-subtracted radial profiles. The weight was calculated as the ratio of the unmasked area to the total area. The resulting weighted average profile is then given by 
\beq
\overline{x}(r) = \frac{\sum w_{\rm i} \, (x_{\rm i}(r) - x_{\rm i, bg})}{\sum w_{\rm i}}  \,\, \left( w_{\rm i} = \frac{A_{\rm i, unmask}}{A_{\rm i, total}} \right), 
\label{eq:xprof}
\eeq
where $x_{\rm i}(r)$ is the X-ray count-rate radial profile of the $i$-th filament, $x_{\rm i,bg}$ is the average signal at 10--20 Mpc from the $i$th filament spine, $A_{\rm i, total}$ is the total area, and $A_{\rm i, unmask}$ is the unmasked area within 20 Mpc from the $i$th filament spine. 

\subsection{Results}
\label{subsec:stack}

\begin{figure*}
    \centering
    \includegraphics[width=0.49\linewidth]{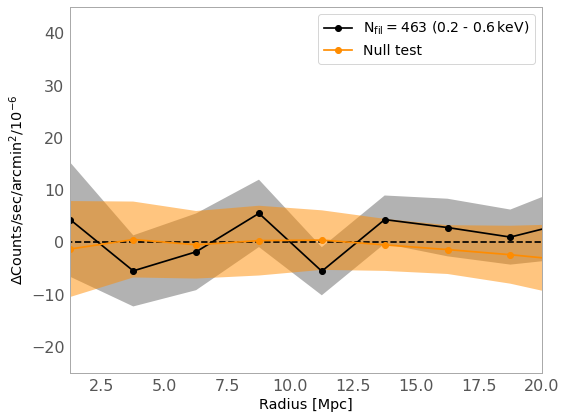} 
    \includegraphics[width=0.49\linewidth]{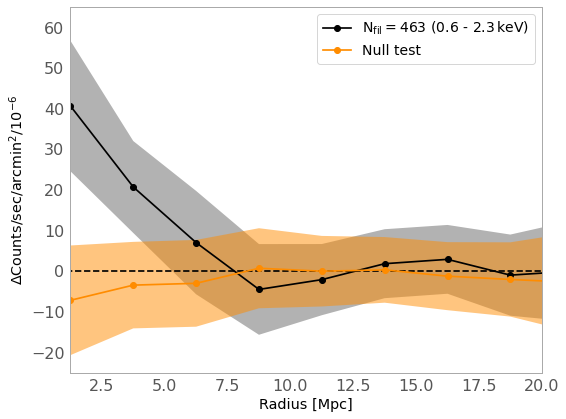} 
    \caption{Average radial X-ray profile of the 463 filaments. The black lines are the ones at 0.2 -- 0.6 keV in the left panel and at 0.6 -- 2.3 keV in the right panel. The 1$\sigma$ uncertainty is estimated by a bootstrap resampling, shown as the gray strip. The average and standard deviation from the 1000 null-test profiles are shown as the orange line and strip, respectively.  }
    \label{fig:xprof}
    \end{figure*}

Following the stacking method described above, we first analyzed the sample of 463 filaments using the eFEDS count-rate map at 0.2 -- 0.6 keV. The resulting stacked radial profile is shown in black in the left panel of Fig.~\ref{fig:xprof}. We then assessed the uncertainty of the profile by bootstrap resampling, also by following the procedure in T20x, in which we drew a random sampling of 463 filaments with replacements and recalculated the stacked profile for the new combination of 463 filaments. We repeated this process 1000 times and produced 1000 stacked profiles. Finally, we computed the standard deviation of this ensemble and used it as the  $1\sigma$ uncertainty of our measurement, which is shown as the gray region in Fig.~\ref{fig:xprof}. As shown in the figure, our measurement at 0.2 -- 0.6 keV is consistent with zero. This is expected from the results of T20x, who stacked the \rosat\ maps at the position of 15\,165 filaments, but did not find a signal excess below 0.56 keV. The reason is likely that at the low-energy bands, the background signal starts to dominate the filament signal, as shown in the predicted spectra in T20x. We next analyzed the sample of 463 filaments using the eFEDS count-rate maps at 0.6 -- 2.3 in the same manner and found an excess signal, as shown in the right panel of Fig.~\ref{fig:xprof}. 

In order to assess the $rms$ fluctuation due to the foreground and background X-ray emissions in the unmasked regions, we performed a null test based on a Monte Carlo method, as conducted in T20x. In the null test, we replaced each of the 463 filaments at random positions on the eFEDS field and then stacked the X-ray count-rate map at the new positions of the 463 filaments. We repeated this null test 1000 times and computed the average and standard deviation of the 1000 null-test set of profiles. 
The result shows that the average of the null profiles is consistent with zero with no discernible structure, as shown in Fig.~\ref{fig:xprof}. This result suggests that our estimator is not biased by the foreground and background X-ray emissions. In addition, the standard deviation of the null profiles is consistent with the uncertainty from the bootstrap estimate (gray area in Fig.~\ref{fig:xprof}), and the ensemble of the null profiles can be used to estimate the uncertainty and significance of our measured signal. 

We hence estimated the S/N of our measured X-ray profile as 
\beq
S/N = \sqrt{\chi^2_{\rm data} - \chi^2_{\rm null}}
\label{eq:snr}
,\eeq
where
\beqa
\chi^2_{\rm data} &=& \sum_{i,j} x_{\rm data}(R_{i})^{T} (C^{-1}_{ij}) \, x_{\rm data}(R_{j}) \\
\chi^2_{\rm null} &=& \sum_{i,j} x_{\rm null}(R_{i})^{T} (C^{-1}_{ij}) \, x_{\rm null}(R_{j}) 
,\eeqa
where $x_{\rm data}(R_{i})$ is the X-ray count-rate at the $R_{i}$ bin from the data X-ray profile,  $x_{\rm null}(R_{i})$ is the value at the $R_{i}$ bin from the null tests, and $C_{ij}$ is the covariance matrix of the data X-ray profile, estimated from bootstrap resampling. The S/N values were estimated to be $\sim$0.9 at 0.2 -- 0.6 keV and $\sim$2.8 at 0.6 -- 2.3 keV using the data points up to 10 Mpc, determined from the extent of the data profile. In addition, the p-values from the $\chi^2$ distribution under the null hypothesis were estimated to be $\sim$0.72 at 0.2 -- 0.6 keV and $\sim$0.012 at 0.6 -- 2.3 keV.
The possible sources of systematic effects are discussed in Sect. \ref{sec:systematics}


\section{Physical interpretation}
\label{sec:interp}

In this section, we perform an X-ray spectral analysis for our measurements based on physical models, and we derive the average gas density and temperature in the filaments. As performed in T20x, our analysis focuses on the core of the filaments because the S/N of our measurements is low at the outskirts, and it does not allow us to determine the physical properties at the outskirts. The size of the filament core was set to $\sim$ 2.5 Mpc based on the temperature profile of filaments from hydrodynamic simulations \citep{Daniela2021}, in which the authors show a flat temperature profile from the center to $\sim$ 2.5 Mpc.

\subsection{Spectral energy distribution}
\label{sec:spec}
To perform our X-ray spectral analysis, we produced new image maps with narrower energy bands using the eROSITA Science Analysis Software System (eSASS) developed by the \erosita\ collaboration. The eSASS includes a collection of tasks, scripts, and libraries to create \erosita\ calibrated science data products and to perform various interactive data analysis tasks  \citep{brunner2022, Predehl2021}. 

As processed in \cite{brunner2022}, using the \textsl{evtool} command in the eSASS software, we merged the four event files into one, applied filters of FLAG = 0xc00fff30 (selecting good events from the nominal field of view, excluding bad pixels), PATTERN $\leq$ 15 (including single, double, triple, and quadruple events), and FLAREGTI (background flare is filtered out), extracted the selected events in specific energy ranges, and created image maps with a resolution of 4 arcsec per pixel. We did not correct the data for the temporary malfunctioning of the camera electronics. Instead, we masked the associated area that included these data. We created image maps in the energy range of 0.4 -- 0.5, 0.5 -- 0.6, 0.6 -- 0.8, 0.8 -- 1.0, 1.0 -- 1.3, and 1.3 -- 2.3 keV. These energy bins were determined to increase the number of energy bands and to avoid that the total number of X-ray photons was low at each energy band. 
We excluded the X-ray events below 0.4 keV and above 2.3 keV because the S/N at these energy bands was consistent with zero, and adding these data did not change our results. From these new image maps, we recomputed the count-rate maps.

Before we proceeded with the stacking analysis for the new count-rate maps, we additionally masked the overlap area of filaments at different redshifts in 0.2<$z$<0.6, as performed in T20x. This masking step avoids accumulating X-ray signals from possible overlapping filaments in the same line of sight and thus gives a conservative X-ray signal for our filament sample. The mask size was 5 arcmin, corresponding to the average angular core size of our filament sample. After the full masking procedure, we repeated our stacking analysis on the new count-rate maps at the six energy bands and computed the X-ray count-rate profiles. From these profiles, we extracted the average X-ray signals within 2.5 Mpc at the cores of the filaments and obtained the spectral energy distribution (SED) of the X-ray signal, which is shown in black in Fig.~\ref{fig:xspec-fit}.
We estimated the S/N of the SED measurement at 0.4 -- 2.3 keV using the covariance matrix from the 463 filaments and found it to be $\sim$3.8, computed from the excesses at the six energy bands. The p-value from the $\chi^2$ distribution under the null hypothesis was estimated to be $\sim$0.0017.

T20x simulated 15,165 X-ray spectra predicted for \srg{} at the cores of the 15,165 filaments with different sets of gas densities and temperatures. The simulations considered different filament lengths, orientation angles, and redshifts from the filament catalog as well as the effective area outside the cluster and point-source mask. The 15,165 filaments contain the 463 filaments we used in this study, thus we extracted the 463 simulated X-ray spectra, corresponding to our sample, with the gas density and temperature estimated using the \rosat\ data, and we computed the S/N of their stacked X-ray emission relative to the background. The obtained forecast S/N is $\sim$ 4.1, which agrees very well with our data analysis result of S/N $\sim$ 3.8.

\subsection{Model fitting}
\label{subsec:model}

As with T20x, we modeled the X-ray SED measurement using the Python X-ray spectral analysis package (PyXspec) interface with the X-ray spectral analysis package (XSPEC) \citep{Arnaud1996}. 
The ancillary response file (ARF) and response matrix file (RMF) were extracted using the \textsl{srctool} command in the eSASS software.
We considered the X-ray emission based on the Astrophysical Plasma Emission Code (APEC) model \citep{Smith2001} with two free parameters: the normalization of surface brightness and temperature. We assumed the gas metallicity to be $\sim$0.1 of the solar value \citep{anders1989} based on the filament analysis with the IllustrisTNG simulations (private communication with \citealt{Daniela2021}). For the photoelectric absorption, we used the \textsl{phabs} model in XSPEC, and the neutral atomic hydrogen (HI) column density in the eFEDS region was estimated to be $\sim3 \times 10^{20} \rm{cm^{-2}}$ with the HI4PI map \citep{hi4pi2016}. We used the median redshift of the 463 filaments, $z\sim0.48$, for the fitting. 

\begin{figure}
    \centering
    \includegraphics[width=\linewidth]{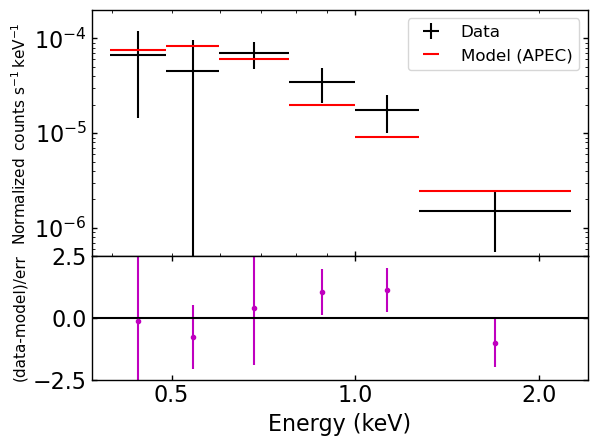} 
    \caption{X-ray spectrum at the core of the 463 filaments. In the top panel, black crosses show excesses of X-ray emission at the core of the 463 filaments (< 2.5 Mpc from the filament spines), relative to the background (average X-ray signal at 10--20 Mpc from the filament spines) at the six energy bands. The line width expresses their uncertainties. The 1$\sigma$ uncertainties are estimated by a bootstrap resampling. Red lines are fit to the data with the APEC model (see text for details). In the bottom panel, the magenta points show the ratio of the data and model at each energy band.}
    \label{fig:xspec-fit}
    \end{figure}

We fit this APEC model to the data with a minimum $\chi^2$ method (see Fig.~\ref{fig:xspec-fit}) and obtained a surface brightness of $(0.04 \pm 0.01) \times 10^{-12} \, \rm erg \, cm^{-2} \, s^{-1} \, deg^{-2}$ at 0.5--2.0 keV and a gas temperature of $1.0^{+0.3}_{-0.2}$ keV. The reduced $\chi^2$ value was 1.0 with a p-value of 0.4. 

The surface brightness is a projected X-ray emission from the gas in filaments and can be converted into a gas density given a density distribution model. While the gas distribution in filaments is not well known, it was studied using the tSZ measurements in T20y, in which the authors showed that a cylindrical filament with a $\beta$-model \citep{Cavaliere1978} with $\beta$=2/3 fit the data well. Thus, we used their model for the gas distribution in filaments. 

The gas distribution in the $\beta$ model is given by 
\beq
n_{\rm e}(r,z) = \frac{n_{\rm e,0}(z)}{1+(r/r_{\rm e,c})^2} \quad \mathrm{in} \,\, r < R_{\rm fil} 
,\eeq
where $n_{\rm e,0}(z)$ is the central electron density of a filament at redshift $z$, $r$ is the radius, and $r_{\rm e,c}$ is the core radius of the electron distribution, and $R_{\rm fil}$ is the cutoff radius. We set $r_{\rm e,c}$ = 1.5 Mpc and $R_{\rm fil}$ = 10 Mpc found in T20y. 
As in T20x, we assumed a negligible evolution of overdensity in filaments in the redshift range of our filament sample at 0.2<$z$<0.6 based on the simulations in \cite{Cautun2014} and used a central electron overdensity $\delta_{\rm e,0}$ as a free parameter. Then, the electron density of a filament at redshift $z$ is given by 
\beq
n_{\rm e,0}(z) = \frac{n_{\rm e,0}(z)}{\overline{n}_{\rm e}(z)} \, \overline{n}_{\rm e}(z) = (1 + \delta_{\rm e,0}) \, \overline{n}_{\rm e}(z=0) \, (1+z)^3, 
\eeq
where $\overline{n}_{\rm e}(z)$ is the mean electron density at redshift $z$ in the Universe. 
We also included an orientation angle of a filament on the plane of the sky in our model because the amplitude of the surface brightness depends on it. The orientation angle of a filament was computed from the 3D position of a filament spine in our filament catalog. With this model, the average gas overdensity at the center of the 463 filaments was estimated to be $21\pm5$. 

We also considered a simple density model: a cylindrical filament with a constant density, given by 
\beq
n_{\rm e}(r,z) = n_{\rm e,0}(z) \quad \mathrm{in} \,\, r < R_{\rm fil} \,\, (\rm constant   \text{ }density \, model), \, 
\eeq
where $R_{\rm fil}$ was set to be 5 Mpc as in T20y.
With this model, the average gas overdensity in the 463 filaments was estimated to be $7\pm2$. 

\section{Systematic effects}
\label{sec:systematics}

\begin{figure}
    \centering
    \includegraphics[width=\linewidth]{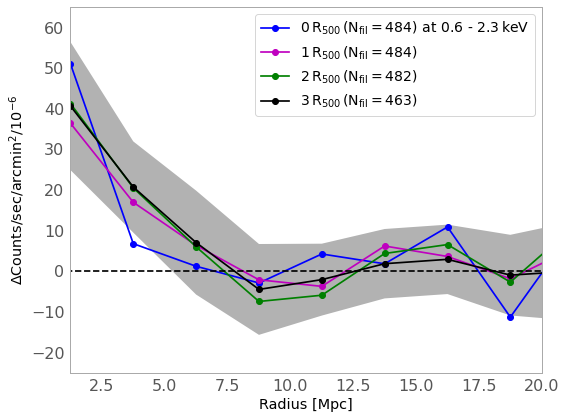} 
    \includegraphics[width=\linewidth]{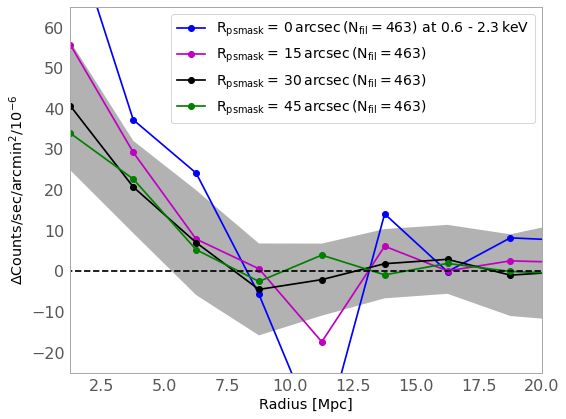} 
    \caption{Contribution to the average X-ray profile of filaments due to galaxy clusters and point sources. Upper: Average radial X-ray profiles of filaments at 0.6 -- 2.3 keV with different sizes for the cluster masks, No cluster mask (blue), $1 \times R_{500}$ (magenta), $2 \times R_{500}$ (green), and $3 \times R_{500}$ (black).
    Lower: Average radial X-ray profiles of filaments at 0.6 -- 2.3 keV with different sizes for the point-source masks. No mask (blue), 15 arcsec (magenta), 30 arcsec (black), and 45 arcsec (green).}
    \label{fig:xprof-mask}
    \end{figure}

Possible bias in our stacked X-ray signals may be caused by residual contamination from foreground or background galaxy clusters. In order to check this, we varied the radius of the circular mask associated with clusters in the range of [0 -- 3] $\times R_{500}$. As demonstrated in the upper panel of Fig.~\ref{fig:xprof-mask}, the result does not show any significant difference in the X-ray profiles when the clusters are masked by more than $1 \times R_{500}$. This shows that our conservative choice of a $3 \times R_{500}$ mask for the clusters makes their contribution to our measured signal quite negligible. 

A bias may also be induced by the contamination from point sources, such as AGN. To check this, we varied the radius of the source mask in the range of [0 -- 45] arcsec. The result is shown in the lower panel of Fig.~\ref{fig:xprof-mask}. We did not see any significant difference in the X-ray profiles when the point sources were masked by a size larger than 30 arcsec, showing that the point sources are well masked with our applied mask size of 30 arcsec. 
In order to further check the residual X-ray emissions outside of the source mask, we stacked the eFEDS point sources at six energy bands and found that $\sim$90\% of the X-ray flux is removed by our 30-arcsecond mask ($\sim$99\% by 45-arcsecond mask), independent of the energy band. 
We note that these residual X-ray emissions take into account all observational effects, such as the point spread function (PSF) and positional uncertainties in the source catalog. 
The average residual X-ray emissions from masked point sources is thus $\sim$10\% with our 30-arcsecond mask. However, this residual flux does not contribute to the signal detected at the position of the stacked filaments, as shown in Fig.~\ref{fig:xprof-mask}, because we applied a background-subtraction procedure. 
To show this, we simulated point-source flux maps at our six energy bands using the stacked point-source images and repeated our stacking analysis of filaments. In the point-source flux map, the point-source flux is painted at the location of the actual point source using the stacked point-source images, but the flux amplitude of each point source is weighted by the count-rate value from the \textsl{ML\_RATE} column in the source catalog. In doing so, we included the actual spatial distribution and relative amplitude of the point sources. For the simulated point-source flux maps at our six energy bands, we repeated our stacking analysis of filaments using the 30-arcsecond mask and found that the X-ray radial profiles are consistent with zero (similar to the left panel in Fig.3). Thus, the Gaussian feature of PSF is removed by the combination of our 30-arcsecond mask and our background-subtraction procedure.
Moreover, to consider the non-Gaussian shape of the PSF, as shown in Fig. B1 of \cite{brunner2022}, we extended the source mask to 60 arcsec and repeated our analysis. 
This 60-arcsecond mask removes $\sim$100\% of the stacked point-source flux as well as the extended wings and spikes in Figs. B1 of \cite{brunner2022}. The results with the 60-arcsecond source mask, summarized in Appendix. \ref{sec:psextend}, were consistent with the results with the 30-arcsecond source mask, suggesting that there is no significant bias in our results that would be due to point sources.

\par\bigskip

The source catalogs used for masking point-sources are either shallow (\rosat{}) or patchy (\xmm{} and \chandra{}). We determined the impact from the shallow and patchy source masks by repeating our analysis without masking the \rosat{}, \xmm{}, and \chandra{} point sources (but masking the eFEDS sources). We found that the impact on our final results is very minor, of about a few percent. This is likely because most of the \rosat{}, \xmm{}, and \chandra{} point sources are included in the eFEDS source catalog and hence are masked by our eFEDS-source mask.

\par\bigskip

Notable straylight features attributed to single-reflected photons from LMC X-1 are shown in Fig. 19 in \cite{Predehl2021}. To investigate the impact of the straylight features, we added a specific 45-arcminunte mask centered on SN1987A to mask out the straylight feature from LMC X-1. After repeating our analysis, we found that its impact is fully negligible. 
No other sources of straylight within the eFEDS field or its vicinity but beyond of the field of view at a given time during the scan path are known. Their contribution to the measured signal should be removed by our background subtraction step. We therefore neglected the contribution of such potential straylight photons in our analysis.

\par\bigskip

The unresolved cosmic X-ray background (CXB) may also contribute to the signal, but this contribution is handled by the background subtraction procedure, which removes local foreground and background signals around each filament. When we included this procedure, we indeed found that the null-test profiles are consistent with zero, indicating that the results are not biased by residual contamination from unresolved AGN and low-mass galaxy groups \citep{Kolodzig2017}. The background-subtraction procedure also removes the diffuse Galactic emission, which is proved by the flat null-test profiles at the low energy band of 0.2 -- 0.6 keV in Fig.~\ref{fig:xprof}.
    
\par\bigskip

We additionally verified whether nonthermal emission could fully explain the detected signals. To to this, we fit our X-ray SED measurements with a power-law model with normalization and photon index as two free parameters. We found the model fits the data not as well, with a reduced $\chi^2$ value of 2.0 (p-value of 0.09) instead of 1.0 with the APEC model. Moreover, in this model, the photon index was found to be $\sim$ 3.1, which is extremely steep compared to the expected value for the AGNs of $\sim 2$ (e.g., \citealt{Ishibashi2010, Yang2015}) and to the one for the CXB of 1.42 -- 1.74 (e.g., \citealt{Lumb2002, Luca2004, Kolodzig2017}). When we fit the data with a power-law model using these realistic photon-index values, the reduced $\chi^2$ value reduces to $\sim$ 2.8 -- 3.2 (a reduced $\chi^2$ value of $\sim$ 2.8 with a p-value of 0.014 for the photon index of 2.0 and $\sim$ 3.2 with a p-value of 0.007 for the photon index of 1.42).
This test suggests that our detected signals cannot be explained by nonthermal emission alone. 
To further check the possible contribution to our detected signal from a nonthermal component, we fit our data with a two-component model: APEC and power law, with the photon index in the range of 1.42 -- 2.0. The result of the fitting shows that the contribution of the power-law component is less than $\sim5 \times 10^{-17} \, \rm erg \, cm^{-2} \, s^{-1} \, deg^{-2}$ (0.5--2.0 keV) compared to $\sim4 \times 10^{-14} \, \rm erg \, cm^{-2} \, s^{-1} \, deg^{-2}$ (0.5--2.0 keV) of the APEC component. 
The reduced $\chi^2$ values were $\sim$1.4 with a p-value of $\sim$0.2 within the photon index of 2.0 and 1.42.
We thus found that the contribution from a nonthermal origin in our measurements is very minor, three orders of magnitudes smaller, relative to that of a thermal origin. 

\par\bigskip

While we assumed a metallicity of 0.1$\zsun$, the metallicity measurements in peripheries of the ICM \citep{Mernier2017, Mernier2018} show the value of $\sim$ 0.2$\zsun$. Therefore, we repeated our spectral analysis with 0.2$\zsun$. Then, the average gas temperature in the filament cores was estimated to be $\sim 1.1$ keV and the surface brightness to be $\sim0.04 \times 10^{-12} \, \rm erg \, cm^{-2} \, s^{-1} \, deg^{-2}$ (0.5--2.0 keV), providing an average gas overdensity at the center of $\sim$18. These values are within the uncertainties of our measurements for 0.1$\zsun$, suggesting that our results contain no significant bias.

\par\bigskip

Finally, we checked whether the fitting of all stacked profiles with a single redshift might impact our final results. To do this, we performed the model fitting with four APEC models with redshifts of 0.25, 0.35, 0.45, and 0.55 by linking their plasma temperatures and normalizations, in which the normalizations were weighted by the relative number of filaments in each redshift bin. The resulting temperature and surface brightness were estimated to be $\sim$0.90 keV and $\sim0.03 \times 10^{-12} \, \rm erg \, cm^{-2} \, s^{-1} \, deg^{-2}$ (0.5--2.0 keV). These values are $\sim$5\% and $\sim$10\% lower than the values estimated with a single redshift, but are still within our measurement uncertainties. 

\section{Discussion}
\label{sec:discussion}

While most of the actual measurements for gas in filaments are still limited to short filaments with lengths smaller than $\sim$ 10 Mpc in the cluster outskirts (e.g., \citealt{Eckert2015, Hattori2017}) or between galaxy clusters (e.g., \citealt{Epps2017, Alvarez2018, Bonjean2018, Tanimura2019a, Tanimura2019b, deGraaff2019, Reiprich2021, Hincks2022}), T20x and T20y reported X-ray and tSZ measurements from the cosmic filaments with lengths of 30--100 Mpc using \rosat\ X-ray data and \planck\ tSZ data, respectively. In their studies, the gas overdensity at the core of filaments was estimated to be $\delta=30\pm15$ in T20x and $\delta = 19.0^{+27.3}_{-12.1}$ in T20y, assuming the $\beta$-model. These values are consistent with our current results of $\delta=21\pm5$. 

\par\bigskip

\cite{Ursino2006} predicted in their simulations that the WHIM contribution to the total diffuse X-ray background might reach up to 20\% in 0.37--0.925 keV, most of which originates from filaments at redshifts between 0.1 and 0.6. The detected signal from our sample of filaments at 0.2 < $z$ < 0.6 peaks at $\sim$ 0.7 keV, as shown in Fig.~\ref{fig:xspec-fit}. This is consistent with these predictions. The gas temperature was estimated to be $0.9^{+1.0}_{-0.6}$ keV in T20x, which is consistent with our result of $1.0^{+0.3}_{-0.2}$ keV. The uncertainty of the temperature estimate is significantly improved by a factor of 2 -- 3 relative to that in T20x. Given the similar level of S/N between these two measurements, this improvement is likely attributable to the improved energy resolution and the extended energy range of \srg{} with respect to \rosat{}.  

\par\bigskip

However, by contrast, the gas temperature estimated in T20y ($0.12 \pm 0.03$ keV) is somewhat lower than our result and than the result in T20x.  The difference may be caused by the fact that the gas temperatures in our study and in T20x are estimated at the filament core (< $\sim$ 2.5 Mpc), whereas the value in T20y is the average within 5 Mpc. 
This temperature difference may imply a radial gradient of the gas temperature: hotter at the core than in the outskirts. Such a radial gradient of the gas temperature in filaments was predicted by hydrodynamic simulations in \cite{Gheller2019}, in which the authors showed that the gas temperature is constant up to $\sim$2.5 Mpc from the filament core with a temperature of $\sim3 \times 10^6$ K ($\sim$ 0.3 keV) and starts to drop beyond this distance. The gas temperature profile in filaments was also predicted with the IllustrisTNG simulations in \cite{Daniela2021}. The authors showed that the WHIM gas temperature profile of different lengths of filaments is constant up to $\sim$2.5 Mpc from the filament core with a temperature of $ (0.8 - 2.0) \times 10^6$ K ($\sim$ 0.05 -- 2 keV), depending on the filament length, and converges to $ \sim 0.9 \times 10^6$ K ($\sim$ 0.01 keV) at 20 Mpc from the filament core. Thus, the gas temperature gradient is predicted in hydrodynamic simulations, but our estimated gas temperature is higher than the predictions from hydrodynamic simulations. The reasons for this difference is still unclear because the uncertainties of our temperature measurements are relatively large, but future \srg\ data will provide an all-sky X-ray measurement that will be helpful in understanding the physical properties of gas in filaments in more detail. 

\section{Conclusions}
\label{sec:conclusion}

We selected 463 filaments that overlap with the eFEDS field at 0.2<$z$<0.6, ranging from 30 Mpc to 100 Mpc in length. They were identified in the SDSS survey. We stacked the eFEDS count-rate maps around these filaments, excluding resolved galaxy groups and clusters as well as the detected X-ray point sources from the \rosat, \chandra, \xmm, and \erosita\ observations. In our stacking analysis, the average signal in the local background region, 10 to 20 Mpc from the filament spine, was subtracted in order to remove  residual X-ray emission and the diffuse Galactic and extragalactic emissions. From the stacking of the filaments, we detected an X-ray excess signal at the core (within 2.5 Mpc) at a significance of $3.8\,\sigma$ in the EDR count-rate maps in six energy bands from 0.4 to 2.3 keV, provided by the \erosita{} collaboration. In parallel, we found that the probability of the measurement under the null hypothesis is $\sim$0.0017.

\par\bigskip

For the physical interpretation of the detected X-ray excess signal, we performed a spectral analysis of the SED measurement. The detected signal cannot be explained by nonthermal emission. When we fit our measurement with a power-law model using realistic photon-index values in the range of 1.4 -- 2.0 for X-ray emissions from AGN or CXB, the reduced $\chi^2$ value is $\sim$ 2.8 -- 3.2 compared to the reduced $\chi^2$ of 1.0 when we fit with the APEC model assuming a thermal origin. Moreover, when we fit our data with a two-component model, APEC plus power law with the photon index in the range 1.4 -- 2.0, the result shows that the contribution of the power-law component is much lower than that of the APEC component. It is lower by about three orders of magnitude. These results support the hypothesis that our detected signal has a thermal origin and cannot be explained by a nonthermal origin. 

\par\bigskip

We fit our measurement assuming the APEC model for the thermal emission. This provided a surface brightness of $(0.04 \pm 0.01) \times 10^{-12} \, \rm erg \, cm^{-2} \, s^{-1} \, deg^{-2}$ at 0.5--2.0 keV and a gas temperature of $1.0^{+0.3}_{-0.2}$ keV with a reduced $\chi^2$ value of 1.0 with a p-value of 0.4. This surface brightness corresponds to a gas overdensity of $21\pm5$ at the center of the filaments, assuming the gas distribution in filaments with a $\beta$-model with $\beta$ = 2/3, or $7\pm2$ assuming a constant gas distribution within 5 Mpc from the filament spines. 

\par\bigskip

The origin of the thermal emission associated with the X-ray signal is still unclear. In our analysis, the known galaxy groups and clusters with masses higher than $\sim 3 \times 10^{13} \, \msun$ were masked. Thus, possible explanations for the thermal emission are either low-mass halos with masses below $\sim 3 \times 10^{13} \, \msun$ or diffuse gas in the filaments. To distinguish between these contributions, the gas distribution around the low-mass halos needs to be understood; however, it is not well constrained so far. For example, the gas fraction in low-mass halos does not agree among different hydrodynamic simulations (e.g., \citealt{Chisari2018, Schneider2019}). The X-ray signal may also be caused by unresolved clumps in the vicinity of the accretion regions of hot halos \citep{Reiprich2021} and/or unresolved accreting sources in low-mass star-forming galaxies.
To interpret the nature of the thermal emission measure in stacked filaments, \srg\ will provide an all-sky X-ray measurement, and the data will help us to understand the gas distribution around low-mass halos. As an another probe for the gas distribution around halos, the kinetic SZ (kSZ) effect from \planck\ and ACT observations can be used, as shown in \cite{Amodeo2021}, \cite{Schaan2021}, \cite{Tanimura2021}, and \cite{Tanimura2022}. The current sensitivity of the kSZ measurements requires a large number of stacked objects to obtain the gas distribution of the halo, and it does not allow us to study their mass and redshift dependence. However, the upcoming Simons Observatory \citep{SO} and CMB-S4 \citep{CMBS4} with a better sensitivity will allow us to study them and clarify the origin of the thermal emission from the cosmic filaments. 



\begin{acknowledgements}
The authors thank an anonymous referee for the useful comments and suggestions.
This research has been supported by the funding for the Baryon Picture of the Cosmos (ByoPiC) project from the European Research Council (ERC) under the European Union's Horizon 2020 research and innovation programme grant agreement ERC-2015-AdG 695561. The authors acknowledge fruitful discussions with the members of the ByoPiC project (https://byopic.eu/team). 
This work was supported by World Premier International Research Center Initiative (WPI), MEXT, Japan. Kavli IPMU was established by World Premier International Research Center Initiative (WPI), MEXT, Japan. Kavli IPMU is supported by World Premier International Research Center Initiative (WPI), MEXT, Japan. 
This work is based on data from \erosita, the soft X-ray instrument aboard SRG, a joint Russian-German science mission supported by the Russian Space Agency (Roskosmos), in the interests of the Russian Academy of Sciences represented by its Space Research Institute (IKI), and the Deutsches Zentrum für Luft- und Raumfahrt (DLR). The SRG spacecraft was built by Lavochkin Association (NPOL) and its subcontractors, and is operated by NPOL with support from the Max Planck Institute for Extraterrestrial Physics (MPE). The development and construction of the \erosita\ X-ray instrument was led by MPE, with contributions from the Dr. Karl Remeis Observatory Bamberg \& ECAP (FAU Erlangen-Nuernberg), the University of Hamburg Observatory, the Leibniz Institute for Astrophysics Potsdam (AIP), and the Institute for Astronomy and Astrophysics of the University of Tübingen, with the support of DLR and the Max Planck Society. The Argelander Institute for Astronomy of the University of Bonn and the Ludwig Maximilians Universität Munich also participated in the science preparation for \erosita. The \erosita\ data shown here were processed using the eSASS software system developed by the German \erosita\ consortium.

\end{acknowledgements}
\bibliographystyle{aa} 
\bibliography{aanda} 


\appendix
\section{Extended point-source mask}
\label{sec:psextend}

\begin{figure}[h]
    \centering
    \includegraphics[width=\linewidth]{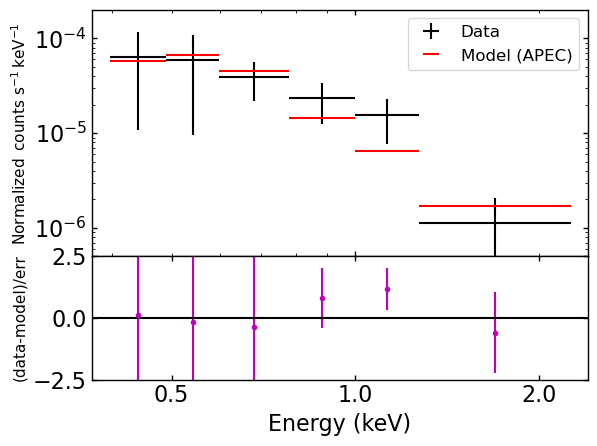} 
    \caption{X-ray spectrum at the core of the 463 filaments using the source mask with the radius of 60 arcsec. In the top panel, black crosses show excesses of X-ray emission at the core of the 463 filaments (< 2.5 Mpc from the filament spines), relative to the background (average X-ray signal at 10--20 Mpc from the filament spines) at the six energy bands when we use the source mask with the radius of 60 arcsec. The line width expresses their uncertainties. The 1$\sigma$ uncertainties are estimated by a bootstrap resampling. Red lines are fit to the data with the APEC model. In the bottom panel, the magenta points show the ratio of the data and model at each energy band.}
    \label{fig:xspec-fit60}
    \end{figure}

The Gaussian contribution of the PSF can be removed by our source mask with a radius of 30 arcsec. However, an actual non-Gaussian shape of the PSF of \srg{}, as shown in Fig. B1 of \cite{brunner2022}, may contaminate our measurements. To avoid the contamination completely, we further extended the source mask to 60 arcsec and repeated our analysis. This 60-arcsecond mask removes $\sim$100\% of the stacked point-source flux, as discussed in Section \ref{sec:systematics}, and the extended wings and spikes in Figs. B1 of \cite{brunner2022}. The SED measurement at our six energy bands with the extended mask is shown in Fig.~\ref{fig:xspec-fit60}, which is consistent with Fig.~\ref{fig:xspec-fit}, which shows the case using the 30-arcsecond source mask. For this SED measurement, we performed the spectral analysis, as done in Section \ref{subsec:model}, and obtained a gas temperature of $\sim 0.9$ keV and a surface brightness of $\sim 0.03 \times 10^{-12} \, \rm erg \, cm^{-2} \, s^{-1} \, deg^{-2}$ at 0.5--2.0 keV, providing a gas overdensity of $\sim 18$. These results are consistent with our derived values with the 30-arcsecond source mask, suggesting that there is no significant bias in our results that would be due to the leakage from the non-Gaussian shape of the PSF.

\end{document}